\documentclass[twocolumn,showpacs,preprintnumbers,amsmath,amssymb]{revtex4}
\usepackage{graphicx}
\begin{document}
\title{ Effect of a columnar defect on the shape of slow-combustion fronts}
\author{M. Myllys, J. Maunuksela,
J. Merikoski, and J. Timonen}
\affiliation{Department of Physics, 
P.O. Box 35, FIN-40014 University of Jyv\"askyl\"a, Finland}

\author{V. K. Horv\'{a}th}
\affiliation{Department of Biological Physics, E\"otv\"os University,
1117 Budapest, Hungary}

\author{M. Ha and M. den Nijs}
\affiliation{Department of Physics, University of Washington,
Seattle, WA 98195, USA}

\begin{abstract}

We report experimental results for the behavior of slow-combustion
fronts in the presence of a columnar defect with excess or
reduced driving, and compare them with those of mean-field
theory. We also compare them with simulation results for an 
analogous problem of driven flow of particles with hard-core 
repulsion (ASEP) and a single defect bond with a different 
hopping probability. The difference in the shape of the front 
profiles for excess vs. reduced driving in the defect, clearly 
demonstrates the existence of a KPZ-type of nonlinear term in the 
effective evolution equation for the slow-combustion fronts. We 
also find that slow-combustion fronts display a faceted form for 
large enough excess driving, and that there is a corresponding 
increase then in the average front speed. This increase in the 
average front speed disappears at a non-zero excess driving in 
agreement with the simulated behavior of the ASEP model.

\end{abstract}
\graphicspath{{figures/}}
\maketitle

\section{Introduction}

Nonequilibrium interfaces that display interesting scaling
properties are quite common in physical (crystal growth, fluid
penetration into porous media, {\it etc.}), chemical (reaction
fronts), as well as biological (growing bacterial colonies)
systems. The dynamics of these systems have long been thought to
be generically described by the Kardar-Parisi-Zhang (KPZ)
equation \cite{KPZ}, or some other equation of motion in the same
universality class \cite{HHZ}. In two space dimensions in
particular (one-dimensional interfaces) when exact solutions are
available, the scaling properties of the KPZ equation are well
understood. The same is not, however, true for the experimental
observations of scaling of interfaces. Typically one has found a
roughening exponent clearly higher than that for the KPZ
equation \cite{HHZ}. Various explanations have been suggested as
for why the KPZ scaling has not generally been found, and most of
the time correlated and/or non-Gaussian noise have been the prime
suspects \cite{HOR,HHZ}.

Recent experiments on slow-combustion fronts propagating in
paper \cite{JKL1,JKL2}, and on flux fronts penetrating a high-$T_c$
thin-film superconductor \cite{AMS}, have provided new insight
into this problem \cite{HOR2}. It indeed appears that
short-range-correlated noise, quenched and dynamical, with
possibly at the same time a non-Gaussian amplitude distribution
for small time differences, induce an additional time and an
additional length scale, beyond which KPZ scaling can only be
observed. Despite these recent advances, it would still be worth
while to demonstrate the existence of the nonlinear term, as
introduced by Kardar, Parisi and Zhang \cite{NEL}, and essential for
the KPZ dynamics, directly from the observed fronts. This would in
essence prove that KPZ type of dynamics, including possibly
effects of nontrivial noise, can indeed be expected to be generic
for nonequilibrium interfaces. One can demonstrate the presence of
this term indirectly by using, {\it e.g.}, inverse schemes able to 
infer the (partial) differential equation that governs the observed
stochastic evolution of interfaces \cite{MAU}, but there is also a 
way to produce a directly observable effect on the shape of the
interface, due to this term.

This method for observing the operation of the nonlinear term was
suggested already for some while ago by Wolf and Tang \cite{WT}. They
considered the effect of columnar defects, columnar in two space
dimensions on which case we concentrate here, and found that there
is a clear 'asymmetry' between the shapes of the fronts that
propagate in the presence of an advancing (excess driving)
and a retarding (reduced driving)
defect. This asymmetry is a direct consequence of the
nonlinear term in the KPZ equation. For a positive coefficient in
this term, applicable to slow-combustion fronts, the noise-averaged 
front should be faceted with a forward-pointing triangular 
shape around an advancing defect, with a height proportional
asymptotically to the width of the sample in the case of a defect
in the middle of the sample, or to the basic period
in the case of periodic boundary conditions. In the case
of a retarding defect, the shape of the front should not be
faceted, and the (negative) height of the deformation in the profile 
should be proportional, according to this mean-field theory, to the 
logarithm of the basic period. Despite its apparent simplicity, 
this kind of experiment has never been performed. 

As is well known, the BCSOS interface model in which the 
nearest-neighbor heights are restricted to differ only by $\pm 1$, 
displays KPZ behaviour, and is on the other hand equivalent to a 
driven flow of particles (hopping rate $p$) with hard-core 
repulsive interactions (ASEP) \cite{HHZ}. A columnar defect in 
an interface model corresponds to a fixed slow or fast bond 
(hopping rate $rp$ with $r<1$ or $r>1$, respectively) in the ASEP 
model. A faceted interface corresponds to a traffic jam of 
infinite length in the thermodynamic limit behind the slow bond. 
Another related question is the detailed shape of the
density/interface profile. The mean-field theory of \cite{WT}
predicts an infinite queue for all $r<1$ and no queue (a logarithmic
decay of the density profile) for $r>1$, {\it i.e.}, $r_c=1$. 
Janowsky and Lebowitz \cite{JL} considered the totally asymmetric 
TASEP model with a slow bond, but concentrated mainly on the shock 
wave fluctuations far away from the slow bond, and apparently did 
not consider the faceting/queueing transition (they had a phase 
diagram with $r_c = 1$). The same model was also considered by 
Kolomeisky \cite{kolo}, but he did not consider the faceting/queueing 
transition either.

Kandel and Mukamel considered a somewhat different model, which
is supposed to be in the same universality class, and
proposed \cite{KM} that the faceting/queueing transition should 
take place at an $r_c <1$. Their simulation data were not, however, 
conclusive.

In slow-combustion experiments the detailed shape of the front 
profile is difficult to determine, and thereby also the 
disappearance of faceting. Faceting is however related to 
increased front speed, also in the thermodynamic limit, and this
is an easier observable. For possible non-faceted fronts, which
would correspond to $r_c <r<1$, an increased front speed would only 
be a finite-size effect, as also the decreased front speed in the
case of a retarding defect corresponding to $r>1$. Notice that
the effective nonlinear term is positive in the slow-combustion
experiments, while it is negative in the ASEP models. Therefore, 
an advancing (retarding) columnar defect in the first case corresponds 
to a slow (fast) bond in the latter case.

\section{Experimental details}

The equipment we use in slow-combustion experiments has been
described elsewhere \cite{JKL1, JKL2}, so it suffices to say here
that samples were 'burned' in a chamber with controllable
conditions and that the video signal of propagating fronts was
compressed and stored on-line on a computer. The spatial
resolution of the set up was $120\ \mu$m, and the time resolution
was 0.1 s. For the samples we used the lens-paper grade (Whatman) 
we have used also previously \cite{JKL2}. Lens paper was now used 
to speed up the experiments.

As slow-combustion fronts do not propagate in paper without adding
an oxygen source for maintaining the chemical reaction involved,
we added as before \cite{JKL1,JKL2} a small amount of potassium
nitrate in the samples. This method also allows for a relatively 
easy way to produce advancing and retarding columnar defects. By 
using masks it is straightforward to produce a narrow (vertical) 
stripe with a smaller or an additional amount of potassium nitrate. 
The average concentration of potassium nitrate determines the average 
speed of the fronts, so it serves as the control parameter of the 
problem.

It is, however, quite difficult to
accurately regulate the amount of potassium nitrate absorbed in 
the sample. This means that it is difficult to produce samples
with exactly the same base concentration, and the same
concentration difference between the base paper and the columnar
defect. Therefore, the statistics we get for any fixed difference 
in the concentration is not quite as good as we would hope. They 
are adequate for the main features of the fronts but not for such 
details as, {\it e.g.}, accurate forms of the front profiles. They 
are also good enough for a quantitative analysis of changes 
in the front speed.

The samples were typically 20 cm (width) by 40 cm, and the
columnar defect (vertical stripe) in the middle of the sample 
was 1.0 cm wide. The defect cannot in our case be too narrow as
fluctuations in the slow-combustion process would then tend to 
wipe out its effect. Too
wide a stripe would on the other hand cause effects due to its
nonzero width, which are unwarranted. We also used simulations
with a discretized KPZ equation to check that the ratio 1 cm to
20 cm should not cause additional effects \cite{MYL}. The length of
the samples was in most cases adequate for achieving stationary
behavior, and only those results are used here where saturation of
the profile was evident.

When analyzing the front profiles, the stripe was removed from the
data, as well as about 6 mm from both boundaries of the samples.
As the system is symmetric across the stripe in
the middle, the observed front profiles were also symmetrized for
better statistics.

As reported already before \cite{JKL1,JKL2}, fluctuations in the
slow-combustion fronts in paper are noticeable. For extreme values
of the potassium-nitrate concentration there appear problems
with pinning (low concentration) or local avalanche type of bursts
(high concentration) in the fronts. Also, very small values of the
concentration difference between the base paper and the defect
stripe could not be used, as fluctuations then completely masked
the effect of the defect. These problems were noticeable for
retarding defects in particular. In the data reported here,
concentration varied between 0.265 and 0.61 gm$^{-2}$ in the base
paper, between 0.1 and 1.05 gm$^{-2}$ in the stripe, and the
concentration difference varied between 0.06 and 0.49 gm$^{-2}$ on
the positive side (25 burns), and $-$0.197 and $-$0.478 gm$^{-2}$ on 
the negative side (19 burns). Because of the practical restrictions
and fluctuation effects, the number of successful burns was 
relatively small.

\section{Results}

We will need the dependence on potassium-nitrate concentration of
the front velocity below so we consider it first. It is useful to
begin with a discussion of the accuracy of the front-velocity 
determination.

\subsection{Front velocity}

Lens paper is thin so that variations in its mass density and
dynamical effects such as (possibly turbulent) convection around the
combustion front are both expected to give a 
contribution to the effective noise. Noise amplitude is
consequently relatively large \cite{JKL1,JKL2}, and therefore also
velocity distribution of a propagating front can be expected to be
broad. We show in Fig. 1 that distribution for two different
values of the potassium-nitrate concentration.

\begin{figure}[floatfix]
\begin {center}
\includegraphics[width=0.95\columnwidth]{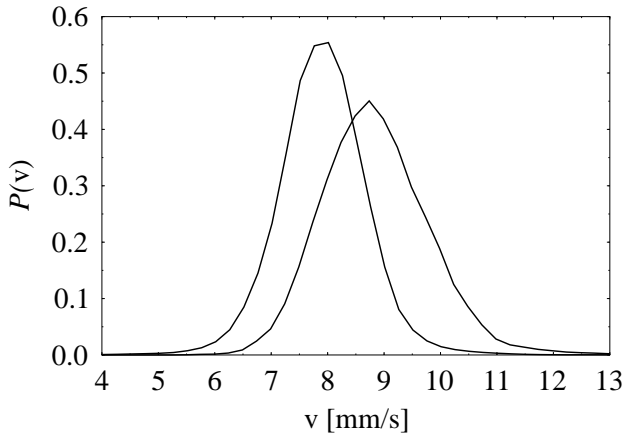} 
\end {center}

\caption{Velocity distributions for slow-combustion fronts with 
potassium-nitrate concentrations 0.34 gm$^{-2}$ (full line), and 
0.536 gm$^{-2}$ (dashed line). Velocities are determined for a 
time difference of 2 s. } \label{fig1}
\end{figure}

It is evident from this figure that, even though the average
velocities in these two cases are relatively well
separated and easily distinguishable,
the velocity distributions have a big overlap.
Together with the limited statistics for any fixed value for
(the difference in) 
the potassium-nitrate concentration, these broad distributions
mean that some variation can be expected to occur in the
measured average velocities (velocity differences) of the fronts.

In the presence of a columnar defect, we should, {\it e.g.}, determine
the change in the average front speed arising from the defect.
This can be accomplished by analyzing separately the undeformed 'flat'
part of the fronts, and the part of the front profile affected by
the presence of the defect. Determination of the average speed of the
flat fronts is done in the transient (wrt profile shape) phase 
in which the (growing) width of the deformed profile is still less than 
the width of the sample. In this phase the flat part of the front is
already in the saturated regime with constant average velocity.
The average speed of the deformed profile is determined in a later
phase in which the width of the deformed part of the profile
essentially coincides with the sample width.

We have also determined the average front velocity for 122 individual
burns for a fairly broad interval in the potassium-nitrate
concentration, and these data are shown in Fig. 2 together with a
linear fit to the measured points.

\begin{figure}[floatfix]
\begin {center}
\includegraphics[width=0.95\columnwidth]{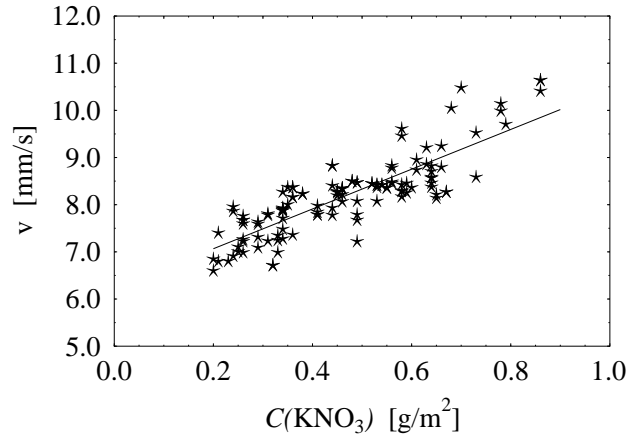}
\end {center}

\caption{Front velocity as a function of potassium-nitrate
concentration. Measured points are denoted by circles and the line
is a linear fit to these points.} \label{fig2}
\end{figure}

The dependence on potassium-nitrate concentration of the front
velocity is not expected to be linear especially near the pinning
limit, but, for the concentration range shown here, it is well
approximated by a linear behavior, which is also more convenient
for the subsequent analysis. We find that, with a linear
fit to the data, the front velocity $v$ is given on the average by

\begin{equation}
v = 4.2C + 6.2, \label{fit}
\end{equation}

\noindent
where $C$ is the potassium-nitrate concentration, and $v$ is in
units mms$^{-1}$ when $C$ is expressed in gm$^{-2}$.

Before showing the measured front profiles in the presence of a
columnar defect, let us consider, in order to make later
comparisons more transparent, what is expected from the mean-field
solution as reported in Ref. \cite{WT}.

\subsection{Mean-field prediction}

We assume that the time evolution of the fronts $h(x,t)$ is
governed by the KPZ equation $\partial h/\partial t =
\nu\nabla^{2}h + {\frac{\lambda}{2}}(\nabla h)^{2} + \kappa + \eta
(x,t)$, where $\eta (x,t)$ describes white noise with
delta-function correlations in space and time, and the driving
term contains the idealized defect as a delta-function
contribution, $\kappa = \kappa_0 + \kappa_1\sum_n\delta (x - L/2
+nL)$. For simplicity we assume here as in \cite{WT} periodic
boundary conditions. If we average over noise in the KPZ equation,
and denote $H(x,t)\equiv \langle h(x,t)\rangle$, we find that

\begin{equation}
\frac{\partial H(x,t)}{\partial t}=\nu
H^{\prime\prime}+\frac{\lambda}{2}(H^{\prime})^2 + \frac{\lambda}{
2}\langle (\nabla\delta h)^2\rangle + \kappa , \label{kpzH}
\end{equation}

\noindent
in which $H^{\prime}\equiv H^{\prime}(x,t)$ denotes the spatial
derivative of $H$, and $\delta h\equiv h(x,t)-H(x,t)$ describes
fluctuations around the noise-averaged profile. A corresponding
equation can be derived for $\delta h$ \cite{WT}.

As $\delta h$ should not depend (locally) on $H(x,t)$ nor on
$H^{\prime}(x,t)$, and only local interdependence between $\delta
h$ and the noise averaged profile can be assumed to appear, one
would then expect \cite{WT} that in leading order 
$\langle (\nabla\delta h)^2\rangle = a_0 + a_{2}H^{\prime\prime}(x,t)$, 
with $a_0$ and $a_2$ some constants. This assumption will make Eq. 
(\ref{kpzH}) closed so that it can be solved without further 
reference to the fluctuations. The delta-function contribution in 
the driving term will induce cusps in $H(x)$ at $x = L/2 - nL$, 
and the solution of Eq. (\ref{kpzH}) is therefore equivalent to 
solving the equation

\begin{equation}
\frac{\partial H(x,t)}{\partial t}=\nu_e
H^{\prime\prime}+\frac{\lambda}{2}(H^{\prime})^2 + \kappa_e
\label{bur}
\end{equation}

\noindent
in the interval $-L/2\leq x\leq L/2$, with boundary conditions
$H(L/2)=H(-L/2)$ and $H^{\prime}(\pm L/2)=\pm s$. Here we have
defined the effective (renormalized) parameters $\nu_e\equiv \nu +
\lambda a_2/2$ and $\kappa_e\equiv\kappa_0+\lambda a_0/2$, and the
magnitude of the slope of the front at the defects is
$s\equiv\kappa_1/2\nu_e$.

Equation (\ref{bur}) is the well-known Burgers
equation \cite{BUR} which can be solved in closed form in one space
dimensions. It is useful to express it first in dimensionless
form, which can be achieved with transformations $H=H_0\tilde H,\
x=H_0\tilde x/s,\ t=2H_0\tilde t/(\lambda s^2)$, with $H_0\equiv
2\nu_e/\lambda$ the internal length scale of the system.

We look for a stationary solution of this equation in the form
$\tilde H(\tilde x,\tilde t) = (\kappa_e +
{\rm sgn}(\kappa_1)q^{2})\tilde t + {\rm ln}(f(q\tilde x))$, where 
${\rm sgn}(z)$ is the sign of $z$, and we have already used the Hopf 
transformation in the spatial part of the ansatz to remove the 
nonlinearity from the equation for $f$. We find that \cite{WT}

\begin{equation}
f(z) = \cosh(z),\ \ q\tanh\left(\frac{q\tilde L}{2}\right) = 1, \label {pos}
\end{equation}

\noindent 
for advancing defects ($\kappa_1>0$), and

\begin{equation}
f(z) = \cos(z),\ \ q\tan\left(\frac{q\tilde L}{2}\right) = 1, \label {neg}
\end{equation}

\noindent
for retarding defects ($\kappa_1<0$). Here $\tilde L\equiv sL/H_0$
is the dimensionless width of the system, and $z\equiv q\tilde x$. 
Asymptotically, for
$\tilde L\gg 1$ (and $\lambda>0$), the profile around an advancing
defect is a forward-pointing triangle with sides that have slopes
$\pm s$, and with height $\Delta H_{+}\equiv H(L/2)-H(0)\simeq
sL/2$. The asymptotic profile around a retarding defect is given
by $H_{0}{\rm ln}|{\rm cos}(\pi x/L)|$ so that $\Delta H_{-}\simeq
-H_{0}{\rm ln}(sL/\pi H_0)$. The magnitude of $H_{+}$ thus grows
linearly with $L$ (or $s$) while that of $H_{-}$ only grows
logarithmically with $L$ (or $s$). This asymmetry is a direct
consequence of the nonlinear term that enhances the deformation in
the former case but reduces it in the latter case.

\subsection{Measured front profiles}

In the above mean-field theory, $\delta v\equiv \kappa_1/L$ is the
difference between the front velocity (driving) inside the defect 
and outside the defect. In the slow-combustion experiments this
velocity difference is regulated by the potassium-nitrate
concentration so that now $\delta v = 4.2\Delta C$, where the
numerical factor comes from the linear fit given by Eq.
(\ref{fit}), and $\Delta C\equiv C_{\rm defect}-C_{\rm base}$ is the
concentration difference. This means that the scaling factor $s$
is given by $s=4.2L|\Delta C|/\nu_e$. Without as yet knowing the
actual value of $\nu_e$ needed for evaluating the size of $s$ and
$H_0$, reasonable estimates, based on the results from the inverse
method solution for the effective equation of motion \cite{MAU},
indicate that the slow-combustion fronts are not necessarily in
the strictly asymptotic regime: we expect that $\tilde L>1$ but
not by a very big margin. Notice that the size of $\tilde L$ is
now regulated by $\Delta C$ as the width $L$ of the samples is
held fixed. Despite the achievable values of $\tilde L$, we can
expect to clearly see the 'asymmetry' in the heights of the front 
profiles for different signs of the concentration difference. In 
Fig. 3 we show the averaged (and symmetrized) front profiles for 
$\Delta C = \pm 0.33$.

\begin{figure}[floatfix]
\begin {center}
\includegraphics[width=0.95\columnwidth]{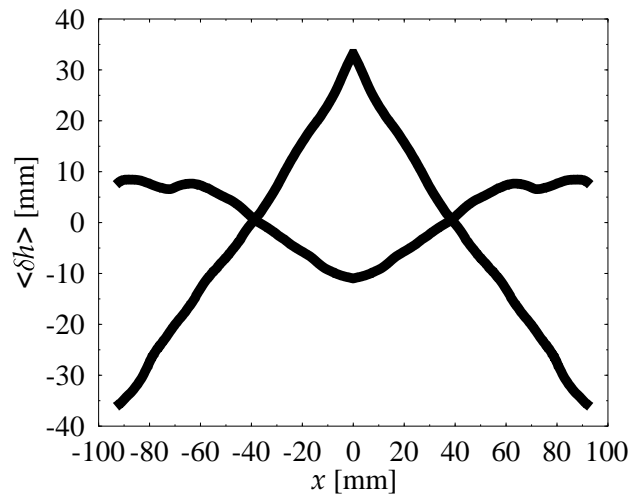} 
\end {center}

\caption{Average front profiles with a columnar defect for $\Delta
C = \pm 0.33$ (upper and lower profile, respectively).}
\label{fig3}
\end{figure}

It is indeed evident that there is a clear difference in the heights
of the front profiles around advancing and retarding defects. By 
following the transient time evolution of the fronts, we could also 
see a clear difference there. For $\Delta C>0$, when a triangular 
deformation was formed after a while around the central stripe, its 
height and base length grew with a more or less constant velocity 
until the base length reached the width of the sample, while the 
slopes of the sides of this triangle remained roughly constant. For 
$\Delta C<0$ on the other hand, the height of the deformation 
saturated much faster even though it also grew more or less linearly 
in time in the beginning, and the base length of the deformation 
reached the sample width at the same time. This transient behavior 
will be analyzed in more detail below. The qualitative behavior for 
the $\Delta C>0$ case is clearly visible in Fig. 4 which shows the
successive fronts with a time difference of 0.5 s for 
$\Delta C=0.327$.

\begin{figure}[floatfix]
\begin {center}
\includegraphics[width=0.95\columnwidth]{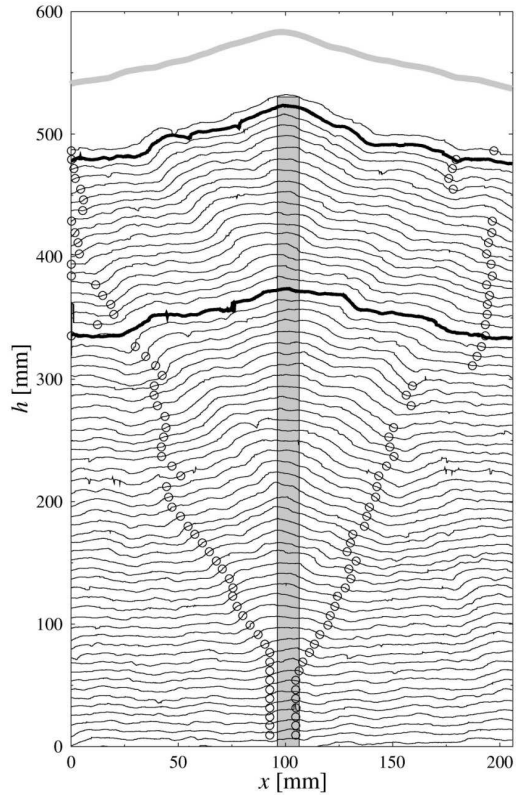}
\end {center}

\caption{Successive fronts with a time difference of 0.5 s for 
concentration difference $\Delta C=0.327$. Also marked are the
stripe with the excess concentration of potassium nitrate, the
fronts between which the average profile is determined (thick
lines), the height of the final profile, and the average shape of
the profile.} \label{fig4}
\end{figure}

A more quantitative comparison between the mean-field
solution for the noise averaged front and the observed
slow-combustion fronts can also be made. For this purpose we have
found it convenient to consider instead of the profile heights
$\Delta H_{\pm}$ the average slopes of the left-hand (LH) sides of
the profiles ({\it c.f.} Fig. 3), $k_{\pm}\equiv 2\Delta
H_{\pm}/L$. As we do not expect to be in the strictly asymptotic
regime, we have used the full transcendental equations for q in
Eqs. (\ref{pos}) and (\ref{neg}) above when fitting the observed
$k_{\pm}$ with the mean-field result.

The average slopes, as functions of concentration difference 
$\Delta C$, will now depend on two parameters, $A\equiv H_0/L$ 
and $B\equiv 2.1L/\nu_e$, which  are used to fit the measured 
slopes. From the fitted values for these parameters we can then 
estimate the coefficients $\lambda$ and $\nu_e$ for this system.

We show in Fig. 5 the experimentally determined values for $k_{+}$
and $k_{-}$ together with the fit by the mean-field solution using
Eqs. (\ref{pos}) and (\ref{neg}).

\begin{figure}[floatfix]
\begin {center}
\includegraphics[width=0.95\columnwidth]{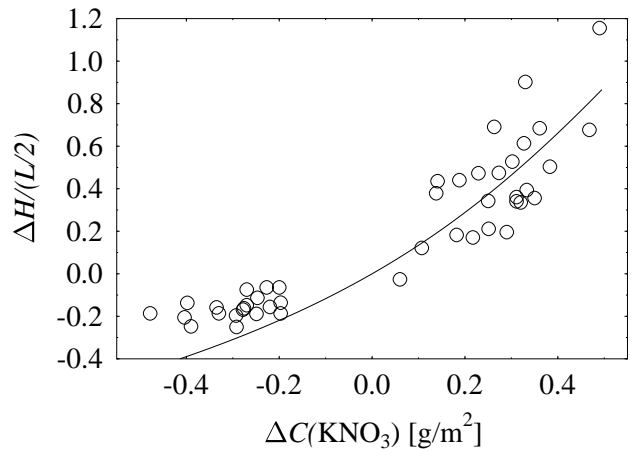}
\end {center}

\caption{The absolute value of the average LH slope of the
deformed front profile around a columnar defect as a function of
difference in the potassium-nitrate concentration. The full line
is a fit by the mean-field solution with Eqs. (\ref{pos}) and 
(\ref{neg}).} \label{fig5}
\end{figure}

Fits to the data were not very sensitive to the actual value of
parameter $A$ so that the correlation coefficient did not change
much even if $A$ was changed in a relatively large interval. If the
$\Delta C>0$ and $\Delta C<0$ data were fitted separately 
without any restrictions on the two parameters, these fits had also 
a tendency to produce somewhat different values for the two cases. 
As the signal-to-noise ratio is better for the $\Delta C>0$ 
data, we fixed $A$ such that it was between the two separately
fitted values but closer to the one from the 
unrestricted two-parameter fit to the $\Delta C>0$ data, and in the 
interval within which the quality of this fit was essentially 
unchanged: $A\simeq 0.3$. Thereafter an unrestricted one-parameter 
fit to the whole data was used to find the value for $B$. In this 
way we found that $B\simeq 2.5$.

The fitted values for parameters $A$ and $B$ allow now an estimation 
of two physical parameters, the 'renormalized' diffusion coefficient
$\nu_e$ and the coefficient of the nonlinear term, $\lambda$. We thus 
find that $\nu_e\simeq 144$ mm$^2$s$^{-1}$, and 
$\lambda\simeq 5.6$ mms$^{-1}$. In the estimate for $\nu_e$ we used 
an 'effective' sample width $L_{\rm eff}\simeq 180$ mm, a bit smaller 
than the width 202 mm of the actual sample, due to the width of the 
defect stripe and to allowing for some boundary effects. By other 
methods we have found previously that $\lambda\simeq 4.1 - 5.1$ 
mms$^{-1}$ \cite{MAU}, so that the value found here is fairly close to 
these previous estimates.

In view of the unavoidable fluctuations in the measured
averaged slopes, we find the fits to the measured points by the
mean-field solution to be quite reasonable.

\subsection{Defect-induced change in front velocity and the queueing transition}

As already discussed above, the mean-field solution predicts a
faceting or queueing transition at $\Delta C = 0$. Above this
transition ($\Delta C>0$), the average front velocity is increased 
due to the presence of an advancing columnar defect, and below this 
transition the change in front velocity should 
vanish for large enough $\tilde L$. For negative $\Delta C$ the 
change in velocity is negative, and should decrease in magnitude 
with increasing $|\Delta C|$. According to
Kandel and Mukamel \cite{KM}, this transition should appear 
at a $\Delta C_{\rm cr}>0$.

As the numerical data of \cite{KM} is not decisive, we have done 
\cite{Ha} simulations on a totally asymmetric ASEP model with a 
fixed defect bond with hopping rate $rp$ in the middle of the
system, while the hopping rate 
at the other bonds was $p$. Open boundary conditions were imposed 
such that the hopping-in rate at the left boundary was $\alpha p$, 
and the hopping-out rate at the right boundary was $\beta p$.
In what follows we only consider the case $\alpha = \beta = p = 1/2$.

This model shows \cite{Ha} a queueing transition at 
$r = r_c = 0.80\pm 0.02$.
In addition, the density profile displays a qualitatively similar 
asymmetry between the slow and fast defect-bond cases as the
mean-field solution for the KPZ fronts between the 
advancing and retarding columnar-defect cases. In the thermodynamic 
limit the deformation stays non-zero only in the faceted phase above 
the transition. The density profiles in both the faceted and 
non-faceted phases also display interesting power-law tails.

As the detailed shapes of the front profiles are difficult to
determine experimentally, we only compare the results for the 
dependence of the average front velocity $V\equiv\langle v\rangle$ 
(current $J\equiv\langle j\rangle$) on the potassium nitrate 
concentration $C$ (hopping rate $p$). In this comparison
dimensionless variables are used, $(V-V_0)/V_0$ for the change in 
the front velocity, and similarly for the current but for reversed sign
as an advancing defect corresponds to a slow bond, $\Delta C/C_0$ for 
the potassium-nitrate concentration difference, and $\Delta p/p=1-r$ for 
the hopping-rate difference. Differences are all determined between the
value with or at the defect and the value elsewhere or without the 
defect. In this way no fitting is involved 
in the comparison. Obviously the actual driving force 
is not known exactly for the slow-combustion fronts, but the observed 
linear dependence well above the pinning transition between the
potassium-nitrate concentration and the front velocity suggests that
the dimensionless difference can be reliably used in this kind of
comparison.

\begin{figure}[floatfix]
\begin {center}
\includegraphics[width=0.95\columnwidth]{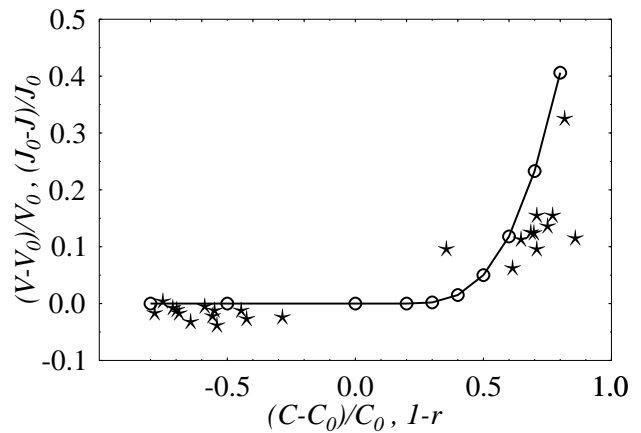}
\end {center}
\caption{Scaled velocity change of slow-combustion fronts due to a 
columnar defect as a function of $(C-C_0)/C_0$ ($\star$), and 
scaled current change in the totally asymmetric ASEP model due to a 
defect bond as a function of $(p-rp)/p=1-r$ ($\circ$). The full line 
connects the latter data points.}
\label{fig6}
\end{figure}

This comparison of the slow-combustion experiment and the totally
asymmetric ASEP model results 
is shown in Fig. 6. It is evident from this figure that agreement between 
the two results is reasonable as there is no fitting involved. There are
still fairly large fluctuations in the experimental data, and it is
not possible to have results for very small values of $\Delta C$
as fluctuations tend to wipe out the whole effect, and the system is
then not in the 'asymptotic regime'. These results indicate, however,
that there indeed is a faceting (queueing) transition at a non-zero
$\Delta C_{\rm cr}$ ($\Delta p_{\rm cr}$, {\it i.e.}, $r_c\not= 1$).

\subsection{Transient behavior}

In addition to the stationary profiles analyzed above, it is also
possible, as already indicated, to study the transient profiles,
{\it i.e.}, how the defect-induced profiles grow at the initial 
phases of the process. The transient behavior of the profile 
around an advancing column is particularly simple. The Burgers 
equation Eq. (\ref{bur}) admits in this case a solution of
exactly the same shape as the stationary solution, which grows
linearly in time until its baseline reaches the width of the
sample. Such a 'self-similar' transient does not exist in the case
of negative $\Delta C$, so analytical results for transient
behavior are then difficult to find. In the non-asymptotic regime
$\tilde L\ll 1$ one can however show that the situation is
symmetric, $\Delta H_{-}\simeq -\Delta H_{+}$. One would thus expect
that, at least in our case when $\tilde L$ is not particularly
large, the height $|\Delta H_{-}|$ would also grow initially (at 
least nearly) linearly in time.

The expected transient behavior for $\Delta C>0$ is already
(qualitatively) evident from Fig. 4 above. More quantitatively the
transient time evolution of the height of the deformed profile can
be analyzed, {\textit e.g.}, by plotting $H(2t)$ against $H(t)$.
For a linear time evolution the former value is twice the latter.
In Fig. 7 we show this plot, averaged over 32 individual burns,
including both signs of $\Delta C$.

\begin{figure}[floatfix]
\begin {center}
\includegraphics[width=0.95\columnwidth]{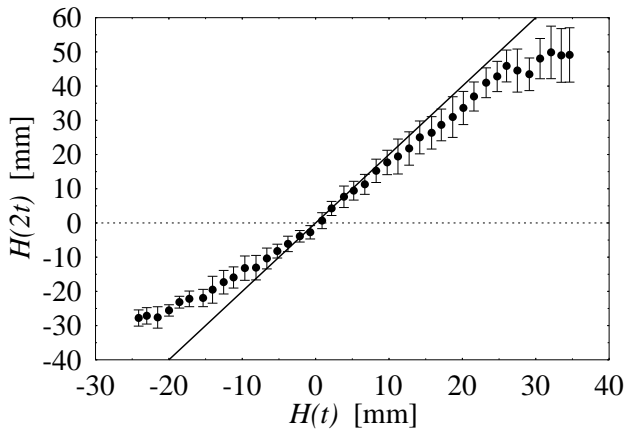}
\end {center}

\caption{$\Delta H(2t)$ as a function of $\Delta H(t)$ averaged
over 32 burns. Positive values correspond to $\Delta C>0$ and
negative values to $\Delta C<0$. The full line is $\Delta H(2t) =
2\Delta H(t)$.} \label{fig7}
\end{figure}

The initial transient behavior is approximately linear in time for 
both cases. For $\Delta C>0$ the trend continues nearly linear until
saturation sets in when the width of the profile equals the width
of the sample. For $\Delta C<0$ the behavior is quite similar
except that saturation takes place earlier. There is also some
indication that, in this case, the growth of $\Delta H$ becomes
nonlinear in time already before saturation, but the quality of
the data does not allow for a decisive conclusion on this.

\section{Discussion and conclusions}

The difference in the amplitude (height), and perhaps not so
clearly in the shape, of the front in the slow-combustion 
experiments, caused by a columnar defect with excess or reduced 
driving, respectively, was clearly demonstrated. The
behavior of the height of the deformed profile, and the 
qualitative shape of the profile in the case of excess driving, 
were also reasonably well explained by the mean-field solution 
of Ref. \cite{WT}. The asymptotic shape of the profile
in the case of negative velocity difference could not be
unequivocally determined as fluctuations are more important in
this case of relatively small amplitude of the profile.
The reduced, in comparison with the case of excess driving,
height of the profile was very evident. In the case of positive
velocity difference the transient behavior of the profile,
{\it i.e.}, the growth of the defect induced deformation in
the profile shape, could as well be explained by the mean-field
solution. For negative velocity difference a nearly linear
behavior in time was observed initially, followed perhaps by a
regime of nonlinear time evolution before saturation.

Fitting the average height (or equivalently the average slopes of
the sides) of the profile with the mean-field solutions 
provided us with estimates for the effective
'diffusion constant' $\nu_e$ and the coefficient of the nonlinear
term, $\lambda$. The latter parameter can also be determined from
the slope dependence of the local front velocity \cite{JKL2,MAU},
or by applying an inverse method on the observed fronts \cite{MAU}.
The value found here for $\lambda$ is fairly close to these previous 
estimates, and we find this level of agreement very reasonable in
view of the rather large fluctuations in the present data.

One should, however, notice that the $\lambda$ measured for a
sample depends on the potassium-nitrate concentration in that
sample, and that the average potassium-nitrate concentration 
was not the same in the samples used in the experiments. We
did not take this variation into account, as it can be assumed to 
give a small effect in comparison with the other experimental 
uncertainties, so that the present estimate represents an 
'average' value.

The effective diffusion coefficient $\nu_e$ contains, in addition
to the bare diffusion coefficient of the original KPZ equation, an
unknown renormalization factor due to noise-induced fluctuations 
around the average front profile. We cannot thus get an estimate 
for the 'bare' diffusion coefficient $\nu$, which can be estimated
by other means \cite{MAU}. However, we can conclude that the 
noise-induced renormalization of $\nu_e$ appears to be sizable.

The position and nature of the faceting (queueing) transition in
interfaces affected by a columnar defect (in the ASEP model by 
a defected bond), has been a long-standing problem. The agreement 
found here between slow-combustion experiments with a columnar 
defect and the related TASEP model results, indicates that this 
transition is indeed at a non-zero value of the respective control 
parameter. No scaling properties of the transition could be analyzed 
at this stage, but the TASEP model results also indicate that this 
transition is continuous. It remains an experimental challenge to 
analyze this transition in more detail.

The authors gratefully acknowledge support by the Academy of
Finland (MaDaMe Programme and Project No. 44875), and fruitful
discussions with David Mukamel and Joachim Krug.



\begin{thebibliography}{99}

\bibitem{KPZ} M. Kardar, G. Parisi, and Y.-C. Zhang,
{\it Phys. Rev. Lett.} {\bf 56}, 889 (1986).

\bibitem{HHZ} For a review, see {\it e.g.} T. Halpin-Healy and Y.-C. Zhang,
{\it Phys. Rep.} {\bf 254}, 215 (1995).

\bibitem{HOR} V. K. Horv\'{a}th, F. Family, and T. Vicsek,
{\it Phys. Rev. Lett.} {\bf 67}, 3207 (1991).

\bibitem{JKL1} J. Maunuksela, M. Myllys, O.-P. K\"ahk\"onen, J. Timonen,
N. Provatas, M. J. Alava, and T. Ala-Nissila, {\it Phys. Rev. Lett.}
{\bf 79}, 1515 (1997); M. Myllys, J. Maunuksela, M. J. Alava, T.
Ala-Nissila, and J. Timonen, {\it ibid} {\bf 84}, 1946 (2000).

\bibitem{JKL2} M. Myllys, J. Maunuksela, M. Alava, T. Ala-Nissila,
J. Merikoski, and J. Timonen, {\it Phys. Rev. E} {\bf 64}, 036101
(2001).

\bibitem{AMS} R. Surdeanu, R. J. Wijngaarden, E. Visser, J. M. Huijbregtse,
J. Rector, B. Dam, and R. Griessen, {\it Phys. Rev. Lett.} {\bf
83}, 2054 (1999).

\bibitem{HOR2} For an early similar observation, see V. K. Horv\'{a}th,
F. Family, and T. Vicsek, {\it J. Phys. A} {\bf 24}, L25 (1991).

\bibitem{NEL} See also D. Forster, D. R. Nelson, and M. J. Stephen,
{\it Phys. Rev. A} {\bf 16}, 732 (1977).

\bibitem{JL} S. A. Janowsky and J. L. Lebowitz, {\it Phys. Rev. A}
{\bf 45}, 618 (1992); {\it J. Stat. Phys.} {\bf 77}, 35 (1994).

\bibitem{kolo} A. B. Kolomeisky, {\it J. Phys. A: Math. Gen.}
{\bf 31}, 1153 (1998).

\bibitem{Ha} M. Ha, M. den Nijs, and J. Timonen, unpublished.


\bibitem{MAU} J. Maunuksela, M. Myllys, J. Merikoski, J. Timonen,
T. K\"arkk\"ainen, M. S. Welling, and R. J. Wijngaarden,
{\it Eur. Phys. J.} {\bf B 33}, 193 (2003) .

\bibitem{WT} D. E. Wolf and L.-H. Tang,
{\it Phys. Rev. Lett.} {\bf 65}, 1591 (1990).

\bibitem{KM} D. Kandel and D. Mukamel,
{\it Europhys. Lett.} {\bf 20}, 325 (1992).

\bibitem{MYL} M. Myllys, J. Maunuksela, J. Merikoski, and J. Timonen, 
unpublished.

\bibitem{BUR} J. M. Burgers, {\it The Nonlinear Diffusion Equation}
(Riedel, Boston, 1974) .


\end{thebibliography}
\end{document}